# Anónimos: An LP based Approach for Anonymizing Weighted Social Network Graphs

Sudipto Das    Ömer Eğecioğlu    Amr El Abbadi

Department of Computer Science
University of California, Santa Barbara
Santa Barbara, CA 93106 - 5110, USA
{sudipto, omer, amr}@cs.ucsb.edu

## ABSTRACT

The increasing popularity of social networks has initiated a fertile research area in information extraction and data mining. Anonymization of these social graphs is important to facilitate publishing these data sets for analysis by external entities. Prior work has concentrated mostly on *node identity anonymization* and *structural anonymization*. But with the growing interest in analyzing social networks as a weighted network, *edge weight anonymization* is also gaining importance. We present Anónimos, a *Linear Programming* based technique for anonymization of edge weights that preserves *linear properties* of graphs. Such properties form the foundation of many important graph-theoretic algorithms such as *shortest paths problem, k-nearest neighbors, minimum cost spanning tree*, and *maximizing information spread*. As a proof of concept, we apply Anónimos to the *shortest paths problem* and its extensions, prove the correctness, analyze complexity, and experimentally evaluate it using real social network data sets. Our experiments demonstrate that Anónimos anonymizes the weights, improves $k$-*anonymity* of the weights, and also scrambles the relative ordering of the edges sorted by weights, thereby providing robust and effective anonymization of the sensitive edge-weights. Additionally, we demonstrate the composability of different models generated using Anónimos, a property that allows a single anonymized graph to preserve multiple *linear properties*.

## Categories and Subject Descriptors

E.1 [**Data Structures**]: Graphs and networks; G.1.6 [**Optimization**]: Linear programming; J.4 [**Social and Behavioral Sciences**]: Sociology

## 1. INTRODUCTION

Social networking sites such as MySpace, Facebook, Twitter, and Orkut have millions of registered users, and the resulting social graph structures have millions of vertices (users or social actors) and edges (social associations). Recent research has explored these social networks for understanding their structure [1,4,22], criminal intelligence [24], information discovery [2], advertising and marketing [13], and others [11]. As a result, companies (such as Facebook) hosting the data are interested in publishing portions of the graphs so that independent entities can mine the wealth of information contained in these social graphs. Anonymization of these graphs is paramount to avoid privacy breaches [3, 15]. Consequently, there has also been considerable interest in the anonymization of graph structured data [5–7, 12, 19, 32, 33]. But most of the existing research on anonymization techniques tends to focus on *unweighted* graphs for *node* and *structural anonymization*.

Recent research has shown applications of the *weighted network model* where social graphs are viewed as weighted networks. Examples include analyzing the *formation of communities* within the network [17], *viral and targeted marketing and advertising* [13], *modeling the structure and dynamics* such as opinion formation [28], and analysis of the network for *maximizing the spread of information* through the social links [14], in addition to the traditional applications such as *shortest paths* [9], *spanning trees* [16] etc. The semantics of the edge weights depend on the application (such as users in a social network assigning weights based on "degree of friendship", "trustworthiness", "behavior", etc.), or the property being modeled [17, 28]). For example, consider the **"Los Angeles"** community in Facebook. If we consider that edge weights are inverse of "trustworthiness" (smaller weights correspond to higher trust in the relation), then the $k$ Nearest Neighbors (kNN) query at a particular vertex returns the $k$ most trusted users associated to the queried user, and the *single source shortest paths tree* provides the most trusted paths within the community which might be used for communicating while minimizing chances of a leak. We focus on the problem of anonymization of edge weights in a social graph.

**Edge-weight anonymization: why do we care?** *First*, in many social networks (such as academic social networks [27]), node identity and link structure is public knowledge. But the edge weights, such as "trustworthiness" of user $A$ according to user $B$, is private information. Therefore, for publishing the graph, anonymization of the edge weights is critical, while node identity anonymization might not be needed. *Second*, even in the case where the node identities are anonymized, edge weight anonymization is still important since if an adversary re-identifies a node in the anonymized graph, even more information will be revealed if edge weights are not anonymized.

**Privacy preserving modeling.** For a weighted network model, the magnitude of the weights, distinguishability of weights, as well as relative ordering of edges by their weights is sensitive information. The goal of anonymization is to prevent rediscovering of this sensitive information with a high confidence. For instance, given an edge's weight ($w'$) in the anonymized graph, an attacker must not be able to determine, with high confidence, the weight of that edge ($w$) in the original graph. Similarly, given two edges ($u_1, v_1$) and ($u_2, v_2$) such that their weights in the anonymized graph have the relation $w'[u_1, v_1] < w'[u_2, v_2]$, an attacker must not be able to discern, with high confidence, the relative ordering of the edge weights in the original graph.

Our solution to the problem of edge weight anonymization is to model the weighted graph based on the property to be preserved, and then reassign edge weights satisfying the model to obtain the anonymized graph. We observe that a large class of graph properties can be expressed as *linear properties*:

DEFINITION 1.1. *A **linear property** of a graph is a property expressible in terms of inequalities involving linear combinations of edge weights.*

*Linear properties* form the basis for a gamut of important graph theoretic properties such as *shortest paths, information spread, credentialing in social media, collective actions, minimum weight matching*, etc. Given a directed graph with non-negative edge-weights, our goal is to model such *linear properties* using a collection of *linear inequalities* and *to preserve the structure of the graph as well as the desired linear properties so that the anonymized graph is at least as useful as the original graph in terms of the property being preserved*. Stated formally, the objective of privacy preserving modeling is:

OBJECTIVE 1.1. *To construct a model that **correctly** captures the inequalities that must be obeyed by the edge weights for the **linear property** being modeled to be preserved. Any solution to such a model would ensure anonymization of edge weights, while preserving the linear property under consideration.*

Once the model has been computed, the set of inequalities in the model need to be solved so that the solution can be used to anonymize the weights while preserving the property being modeled. Linear Programming (LP) is a powerful technique for solving such system of linear equalities. We therefore propose Anónimos, a technique for modeling linear properties using a system of linear inequalities and formulate it as an LP problem. This formulation allows us to use off-the-shelf LP solvers for solving the models and anonymizing the graphs. We also show the **composability** property of Anónimos for preserving multiple linear properties in a single anonymized graph. As a proof-of-concept, we consider the *shortest paths problem* since it is a problem of great interest in weighted graphs. The notion of shortest paths is widely applicable in people's use of social media. Most notably, the determination of (or reliance on) shortest paths is critical in the timely transfer of information from one person to another. Applications include messages regarding everything: from social information (e.g., updates about plans among friends in a social network) to frivolous information (e.g., sharing jokes and entertainment information) to information that is potentially serious and consequential (e.g., information about crises or the spread of disease). Moreover, underlying a variety of online relationships is the notion of trust, which is a fundamental concern in e-commerce transactions, personal relationships, and the consumption of news information, among other things. By involving the fewest people in the transfer of information across a network, potential trust violations (e.g., leaks, misinformation from less well-known sources, or security breaches) are minimized. In these and other ways, the retention of information about shortest paths is of broad appeal. Furthermore, it is also useful in modeling other properties such as *kNN* and community formation within complex network models.

**Contributions.**

- We propose Anónimos, a technique for edge weight anonymization of graph structured data that preserves *linear properties* by expressing them as a system of inequalities formulated as an LP problem (Section 2).

**Table 1: Notational Conventions.**

| | |
|---|---|
| $G = (V, E, W)$ | Weighted graph to be anonymized |
| $G' = (V, E, W')$ | Anonymized graph, $W'$ satisfies the model |
| $n, \|V\|$ | Number of vertices in the graph |
| $d$ | Average degree of the vertices of the graph |
| $w[u, v]$ and $w'[u, v]$ | Weight of edge $(u, v)$ in $G$ and in $G'$ |
| $P[u, v]$ | Path from vertex $u$ to $v$ in the graph $G$ |
| $D[u, v]$ | Cost of $P[u, v]$, $\sum_{(u', v') \in P[u, v]} w[u', v']$ |
| $\Pi[v]$ | Predecessor of $v$ in the shortest paths tree |
| $T_i$ | Shortest paths tree with $v_i$ as the source |
| $x_1, \ldots, x_m$ | Variables representing edge-weights |
| $x_{(u,v)}$ | Variable corresponding to edge $(u, v) \in E$ |
| $f(u, \ldots, v)$ | $\sum_{(u', v') \in P[u, v]} x_{(u', v')}$ |
| $\mu$ | Indistinguishability threshold for *k-anonymity* |
| $N_u$ | Edge neighbor of a vertex |

- We use Anónimos to develop models for different variants of the *shortest paths problem* (Section 4). We also demonstrate the composability of the models by composing the models of the *single source shortest paths trees* to construct the model for *all pairs shortest paths*. Anónimos therefore has the ability to preserve multiple linear properties in a single anonymized graph. We further optimize the models (Section 5) that considerably reduces the complexity of the models.

- We prove the correctness of the proposed models, provide a thorough analysis of the complexity of the proposed models, and present the results of experiments (Section 7) on real social network graphs that validate this analysis, while confirming that the anonymity of the sensitive information is preserved.

## 2. ANÓNIMOS IN ABSTRACT

We now introduce Anónimos, and use Kruskal's algorithm [16] for *minimum spanning tree (MST)* as an example. The goal of this technique is to capture the dynamic behavior of the algorithm using a system of linear inequalities. Given the original weighted directed graph $G = (V, E, W)$ with positive edge weights represented by variables $x_1, x_2, \ldots, x_m$ (where each $x_i$ corresponds to an edge $i = (u, v) \in E$; refer to Table 1 for notational conventions), we model the system of linear inequalities in terms of these variables. For example, at every step of Kruskal's algorithm for the MST [16], the edge with the minimum weight amongst the set of remaining edges, and not resulting in a cycle is added to the MST. Let $(u_i, v_i)$ be the edge selected at the $i^{th}$ iteration, and $(u_{i+1}, v_{i+1})$ be the edge selected in the $(i+1)^{th}$ iteration. This implies that $w[u_i, v_i] \leq w[u_{i+1}, v_{i+1}]$. If $x_{(u_i, v_i)}$ and $x_{(u_{i+1}, v_{i+1})}$ are the variables representing these edges in the model, then this outcome is modeled by the inequality $x_{(u_i, v_i)} \leq x_{(u_{i+1}, v_{i+1})}$. Therefore, for every pair of edges $(u_i, v_i)$ and $(u_{i+1}, v_{i+1})$ selected in consecutive iterations, the inequality $x_{(u_i, v_i)} \leq x_{(u_{i+1}, v_{i+1})}$ can be added to the model whenever the given weights satisfy $w[u_i, v_i] \leq w[u_{i+1}, v_{i+1}]$.

Decisions made at each step of the algorithm can similarly be expressed as linear inequalities involving the edge-weights. Thus, the execution of the algorithm processing the graph can thus be modeled as a set of linear inequalities involving the edge weights

as *variables*:

$$\underbrace{\begin{pmatrix} a_{11} & a_{12} & \cdots & a_{1m} \\ a_{21} & a_{22} & \cdots & a_{2m} \\ \vdots & \vdots & \ddots & \vdots \\ a_{k1} & a_{k2} & \cdots & a_{km} \end{pmatrix}}_{\mathbf{A}} \underbrace{\begin{pmatrix} x_1 \\ x_2 \\ \vdots \\ x_m \end{pmatrix}}_{\mathbf{x}} \leq \underbrace{\begin{pmatrix} b_1 \\ b_2 \\ \vdots \\ b_m \end{pmatrix}}_{\mathbf{b}} \quad (1)$$

Any *feasible* solution to (1), except the original set of weights, can be used to anonymize the weights while ensuring that the properties of the graph remain unchanged with respect to the algorithm being modeled. Linear Programming (LP) provides a flexible and powerful technique for solving such a system of inequalities. Since finding a feasible solution to the model is as hard as finding the optimal solution [23], the model in (1) can be formulated as an LP problem:

$$\text{Minimize (or Maximize)} \quad \mathcal{F} = \mathbf{c^t x}$$
$$\text{subject to} \quad \mathbf{Ax} \leq \mathbf{b}$$

Here $\mathbf{A}$ is an $m \times n$ matrix of coefficients, $\mathbf{b}$ an $m \times 1$ column vector, $\mathbf{c}$ a $1 \times n$ row of coefficients, and $\mathcal{F} = \mathbf{c^t x}$ is a linear objective function. Even though feasibility of (1) is enough, the LP formulation and choice of the objective function $\mathcal{F}$ have a number of advantages which we will discuss in Section 2.1.

Any *linear property* which can be expressed as a function of a linear combination of edge weights can be expressed as an LP problem, and hence Anónimos can be used for modeling a wide variety of *linear properties*. We remark that modeling an algorithm's execution (such as Kruskal's algorithm for MST in this section, and Dijkstra's algorithm for shortest paths in the rest of the paper) is an intuitive way for constructing the model, and the model captures the property of the graph in addition to modeling the algorithm. Once the model has been constructed, any off-the-shelf LP solver package can be used to find a solution to the set of inequalities (constraints), and the solution constitutes the weights of the anonymized graph. The model is said to be **correct** if the property being modeled is preserved across anonymization. The **complexity** of the model is the number of inequalities necessary to define the model.

## 2.1 Properties of Anónimos

Our formulation of (1) as an LP problem lends Anónimos some additional properties which are discussed below.

### 2.1.1 Flexibility and Robustness

The LP formulation of (1) allows the variables (representing the edge-weights) to be assigned varying lower and upper bounds to attain different scalings as well as shifts in the relative magnitudes of the solution. Therefore, the publisher of the graph can publish different anonymized versions of the same graph where the edge weights in each published version is different. Additionally, the LP formulation provides *flexibility of choice of objective function*. Different objective functions $\mathcal{F}$ can be used to generate different solution sets, and hence different anonymized graphs can be generated by changing the vector $\mathbf{c}$ without any need for regenerating the model. In addition, if the publisher of the graph wants to minimize the sum of edge weights in the anonymized graph, then $\mathbf{c}$ can be set as a unit vector. Furthermore, if it is required that some weights in the anonymized graph be smaller than other weights, then $\mathbf{c}$ can be chosen with larger coefficients corresponding to these variables, and smaller coefficients corresponding to the rest. These are just some of the examples of flexibility that the objective function $\mathcal{F}$ lends to this problem and the publisher of the graph. Since *any feasible* solution can be used for anonymization, the choice of $\mathcal{F}$ is a *free parameter* and a choice of the publisher of the graph.

### 2.1.2 Composability

Another important property of Anónimos aided by the LP formulation is *composability*, i.e. the ability to combine models representing different linear properties so that *multiple* properties are preserved in a single anonymized graph.

THEOREM 2.1. *The composition of LP models developed as extensions of Anónimos for modeling different* linear properties *do not lead to contradictory constraints, and hence the models are* composable.

PROOF. **Proof by Contradiction.** Let $G = (V, E, W)$ be the original weighted graph. Let $M_1$ and $M_2$ represent models representing *linear properties* $P_1$ and $P_2$ (shortest paths and minimum spanning tree for instance), and let $S_1$ and $S_2$ denote the set of inequalities representing the models. Let us assume that there exist a contradictory pair of constraints in $S_1 \cup S_2$, i.e., there does not exist a single solution for the set of constraints $S_1 \cup S_2$. Since the set $S_1$ is built based on the original set of weights $W$, hence $W$ is a valid solution satisfying $S_1$ (and there are possibly many more solutions satisfying $S_1$). Using similar arguments, $W$ also satisfies $S_2$. Since $W$ satisfies $S_1$ and $S_2$ individually, hence $W$ must also satisfy $S_1 \cup S_2$. This leads to a contradiction that $S_1 \cup S_2$ had a contradicting pair of inequalities. □

Therefore, if a model is developed to preserve the *minimum spanning tree*, and another model is developed for preserving the *shortest paths*, both the models can be *composed* so that the resulting model captures both properties, and a solution to the composed model results in an anonymized graph that preserves both these properties.

## 3. PRIVACY METRICS

**Attack Model.** The goal of edge-weight anonymization is to prevent an adversary from determining with high confidence "sensitive information" corresponding to the edges:

DEFINITION 3.1. *Sensitivity of Weights: The following information about an edge is considered as* sensitive *and should be preserved across anonymization:*

- *The* magnitude *of the weights associated with the edges.*
- Indistinguishability *of the weight of an edge when compared to the weights of other edges.*
- Relative ordering *of the edge weights in the original graph.*

The reason for indistinguishability is obvious — a distinguishable weight would aid re-identification of the edge and possibly its weight. Ordering of weights is sensitive for certain semantics of edge-weights. For instance, in the *"Los Angeles"* community example with weights representing "trustworthiness", if $A$ rates $B$ as more trustworthy compared to $C$, then $w[A, B] > w[A, C]$. Evidently, this ordering is "sensitive" for all the involved users, and an adversary analyzing the anonymized graph should not be able to determine this order with high confidence. In summary, anonymization should not leak sensitive information such that if an attacker re-identifies an edge in the anonymized graph, s/he can neither determine the edge's original weight, nor the ordering of the edge's weight compared to weights of other edges.

Note that the ordering or indistinguishability of edges is particularly interesting in a *neighborhood* of an edge in the graph. For instance, the ordering of $w[u_1, v_1]$ and $w[u_2, v_2]$ is not important if $u_1, v_1, u_2,$ and $v_2$ are not related in the graph. We therefore define an *edge neighborhood* of a vertex (which also captures the relationship amongst the edges) where ordering and indistinguishability is important.

DEFINITION 3.2. **Edge neighborhood of a vertex.** *The edge neighborhood of a vertex $u$, denoted as $N_u$, is the set of edges emanating from the vertex $u$, i.e., edges with $u$ as the source.*

The notion of neighborhood can change depending on the semantics of the application. To address privacy concerns, we use two well known metrics used in data privacy and statistics which will be evaluated in a particular *neighborhood*:

**k-anonymity.** *k-anonymity* [26] is a well known metric used in data privacy for dealing with the *indistinguishability* of data values in an anonymized data sets. We use the following definition of *k-anonymity* in the context of edge weight anonymization:

DEFINITION 3.3. *An edge $(u, v)$ is **k-anonymous** if there exists $k − 1$ other edges $(u, v_i)$ in the neighborhood $N_u$ such that $\|w[u, v] − w[u, v_i]\| \leq \mu$, where $\mu$ is the indistinguishability threshold, i.e., the difference of weights below which two edge weights cannot be distinguished.*

Anónimos can ensure preservation of *k-anonymity* by adding additional constraints of the form $x(u, v) − x(u, v_i) \leq \mu \ \forall (u, v)$ and $(u, v_i)$ in a neighborhood; though at the cost of increasing the *complexity* of the model.

**Spearman rank correlation coefficient.** The Spearman rank coefficient [25], denoted by $\rho$, is a statistical measure of the correlation of ranks or orders of two ranked data sets, and is used to evaluate the *order sensitivity anonymization*. Consider two $n$-tuples $X$ and $Y$, where $X$ corresponds to the edge weights in the given graph, and $Y$ to the edge weights in the anonymized graph. Let the corresponding rank sequences be $x$ and $y$.

DEFINITION 3.4. *Given two ranked data sets $X$ and $Y$, $\rho$ is computed as:*

$$\rho = 1 - \frac{6 \sum d_i^2}{n(n^2 - 1)}$$

*where $d_i = x_i − y_i$ is the difference between the ranks of the corresponding values $X_i$ and $Y_i$, and $n$ is the number of items in each data set.*

The value of $\rho$ lies between $-1$ and $1$; $\rho = 1$ implies perfect correlation, $\rho = -1$ implies perfect negative correlation, and $\rho = 0$ implies no correlation between the two orders. Therefore, $\rho$ gives a measure of anonymity in terms of scrambling the rank-ordering of edge weights in a neighborhood; given a list of edges in the *edge neighborhood* of a vertex, a value of $\rho$ closer to 0 is desirable. In case there are tied ranks, then a somewhat more complicated formula is used. In this case Spearman rank correlation coefficient in terms of the computed ranks $x_i$ and $y_i$ is

$$\rho = \frac{n \sum x_i y_i - \sum x_i \sum y_i}{\sqrt{n \sum x_i^2 - (\sum x_i)^2} \sqrt{n \sum y_i^2 - (\sum y_i)^2}}$$

---

**Algorithm 4.1** Dijkstra'a Algorithm: Shortest paths tree

1: $D \leftarrow (\infty)$ /* Cost of best known path from source. */
2: $\Pi \leftarrow ()$ /* Predecessor in shortest path from source. */
3: $Q \leftarrow v_0$ /* Set of unvisited vertices */
4: $S \leftarrow \phi$ /* Vertices to which shortest path is known. */
5: $D[v_0, v_0] \leftarrow 0$
6: **while** $Q \neq \phi$ **do**
7:     $u \leftarrow ExtractMin(Q)$ /* Unvisited vertex with min cost */
8:     $S \leftarrow S \cup \{u\}$
9:     **for** each vertex $v$ such that $(u, v) \in E$ and $v \notin S$ **do**
10:       **if** $D[v_0, v] > D[v_0, u] + w[u, v]$ **then**
11:         $D[v_0, v] \leftarrow D[v_0, u] + w[u, v]$
12:         $\Pi(v) \leftarrow u$ /* Shorter path exists. */
13:       **else**
14:         /* Do Nothing. */
15:       **end if**
16:       **if** $v \notin Q$ **then**
17:         $Q \leftarrow Q \cup \{v\}$
18:       **end if**
19:     **end for**
20: **end while**

## 4. SHORTEST PATHS PROBLEM

In this section, we demonstrate how Anónimos can be used for modeling and preserving the *shortest paths* property of a graph. As pointed out earlier, shortest paths in a graph are important to various graph applications, and we choose the shortest paths problem as a proof-of-concept. We first describe the technique for modeling *single source shortest paths tree* using the Dijkstra's algorithm [9], and then demonstrate the composability of Anónimos by composing the models generated for *single source shortest paths* to construct the model for *all pairs shortest paths* problem.

### 4.1 Single Source Shortest Paths – Linear Model

We now show how Anónimos can be used for modeling the *single source shortest paths tree*. Given a weighted graph $G = (V, E, W)$, and a source vertex $v_0$, a *single source shortest paths tree* is a spanning tree of the graph where the path from the source to any other vertex in the tree is the shortest path between the pair in $G$. This tree is important in a number of applications; for example, if weights are assigned based on inverse of "trustworthiness", then this tree will provide the paths with greatest "trustworthiness" for transferring confidential information from a specific node while minimizing chances of a leak.

The *single source shortest paths tree* problem can have various naïve anonymization schemes. We solve this problem separately since: *First*, the *single source shortest paths tree* problem subsumes the *k-nearest neighbors* query, since given the shortest paths tree from node $v_0$, we can determine the top-$k$ nearest neighbors in increasing order. An Anónimos based approach preserves this additional property which many naïve solutions cannot preserve. *Second*, this model forms the basis for the *all pairs shortest paths* problem and we use this as a stepping stone towards this goal. Our composition of the models for *single source shortest paths tree* to model *all pairs shortest paths* problem demonstrates the composability of the models (Theorem 2.1).

Dijkstra's algorithm [9] is a well known greedy algorithm for *single source shortest paths tree*; Algorithm 4.1 provides an overview. Given a start vertex $v_0$, at every step the algorithm selects the vertex $u$ with the smallest known cost from $v_0$. The algorithm "relaxes" the neighbors of $u$ whose cost from the source has now decreased because of the selection of $u$. Figure 1 shows an illustration of the execution of Dijkstra's algorithm on an example graph, and the resulting tree. For notational conventions refer to Table 1. In particular, $D[u, v]$ is the cost of the path from the vertex $u$ to $v$, and $f(u, v)$ is $\sum_{(u',v') \in P[u,v]} x_{(u',v')}$. In other words $f(u, v)$ is a shorthand for the sum of the variables representing the edges in a given path.

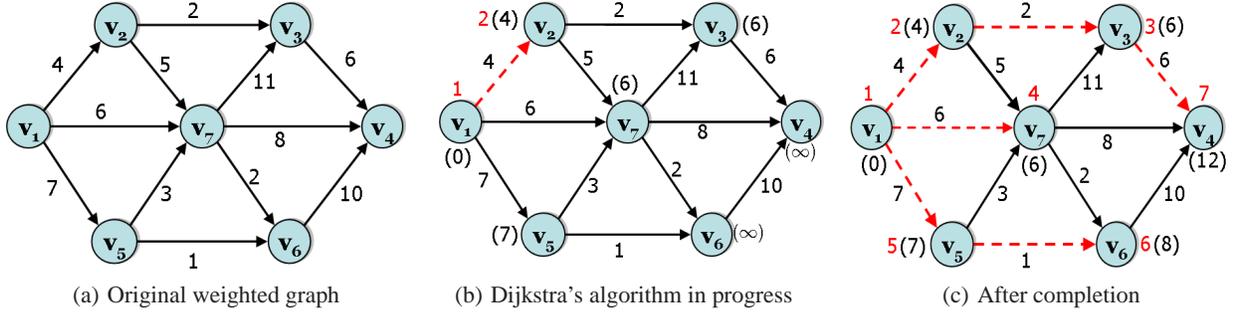

(a) Original weighted graph  (b) Dijkstra's algorithm in progress  (c) After completion

**Figure 1**: Illustration of Dijkstra's algorithm. The numbers adjoining the vertices and outside parenthesis correspond to the order in which the vertices were selected by Dijkstra's algorithm, the number in parentheses correspond to the cost of the best known path from the source, and the dashed edges constitute the shortest paths tree.

Dijkstra's algorithm [9] makes a number of decisions based on the outcome of comparisons of linear combinations of edge weights. These decisions can be modeled using the following three categories of inequalities and are incorporated in Dijkstra's algorithm as shown in Algorithm 4.2:

- **Category I**: When processing edge $(u, v)$, if $D[v_0, v]$ can be improved, then $D[v_0, v] > D[v_0, u] + w[u, v]$, add constraint $f(v_0, v) > f(v_0, u) + x_{(u,v)}$ (line 18 in Algorithm 4.2).

- **Category II**: When processing edge $(u, v)$, if $D[v_0, v]$ can not be improved, then $D[v_0, v] \leq D[v_0, u] + w[u, v]$, add constraint $f(v_0, v) \leq f(v_0, u) + x_{(u,v)}$ (line 20 in Algorithm 4.2).

- **Category III**: When extracting the edge $u$ for the next iteration, if $u'$ is the previous vertex processed, then $D[v_0, u'] \leq D[v_0, u]$, add constraint $f(v_0, u') \leq f(v_0, u)$. This captures the order in which the vertices are selected (line 11 in Algorithm 4.2).

The following theorem formalizes the correctness of this model.

THEOREM 4.1. *A model built from all the inequalities of Categories I, II, and III combined will correctly model Dijkstra's algorithm, i.e., any solution to the model used to anonymize edge weights in the graph results in the same shortest paths tree in the original as well as the anonymized graph.*

PROOF. **Proof by Contradiction.** Let $G = (V, E, W)$ be the input graph, and $G' = (V, E, W')$ be the anonymized graph. Let $T_0$ be the shortest paths tree starting at vertex $v_0$ in $G$ and $T_0'$ be the corresponding tree in $G'$. By way of contradiction, assume that $T_0$ and $T_0'$ are different. Let $v$ be a vertex where $T_0$ and $T_0'$ differ, and let $u$ be its predecessor in $T_0$, and $u'$ in $T_0'$ such that $u \neq u'$. Since $u$ is the predecessor of $v$ in $T_0$ and since $(u, v)$ and $(u', v) \in E$, we must have:

$$D[v_0, u] + w[u, v] = D[v_0, v] \quad (2)$$
$$\text{and,} \quad D[v_0, u'] + w[u', v] \geq D[v_0, v] \quad (3)$$

**Algorithm 4.2** Linear Complexity model
1: $D \leftarrow (\infty)$ /* Cost of best known path from source. */
2: $\Pi \leftarrow ()$ /* Predecessor in shortest path from source. */
3: $Q \leftarrow v_0$ /* Set of unvisited vertices */
4: $S \leftarrow \phi$ /* Vertices to which shortest path is known. */
5: $D[v_0, v_0] \leftarrow 0$
6: $u' \leftarrow \phi$ /* Stores the vertex processed in previous iteration */
7: **while** $Q \neq \phi$ **do**
8:   $u \leftarrow ExtractMin(Q)$
9:   $S \leftarrow S \cup \{u\}$
10:   **if** $u' \neq \phi$ **then**
11:     $AddConstraint(f(v_0, u') \leq f(v_0, u))$
12:   **end if**
13:   $u' \leftarrow u$
14:   **for** each vertex $v$ such that $(u, v) \in E$ and $v \notin S$ **do**
15:     **if** $D[v_0, v] > D[v_0, u] + w[u, v]$ **then**
16:       $D[v_0, v] \leftarrow D[v_0, u] + w[u, v]$
17:       $\Pi(v) \leftarrow u$
18:       $AddConstraint(f(v_0, v) > f(v_0, u) + x_{(u,v)})$
19:     **else**
20:       $AddConstraint(f(v_0, v) \leq f(v_0, u) + x_{(u,v)})$
21:     **end if**
22:     **if** $v \notin Q$ **then**
23:       $Q \leftarrow Q \cup \{v\}$
24:     **end if**
25:   **end for**
26: **end while**

The model will contain constraints corresponding to properties 2 and 3. Again, as $u'$ is the predecessor of $v$ in $T_0'$, and since $(u, v)$ and $(u', v) \in E$, we have:

$$D'[v_0, u'] + w'[u', v] = D'[v_0, v] \quad (4)$$
$$\text{and,} \quad D'[v_0, u] + w'[u, v] \geq D'[v_0, v] \quad (5)$$

Since $W'$ is a solution of the model, properties 4 and 5 will be satisfied only if $u = u'$, which is a contradiction. □

**Complexity of the Model.** Category I and Category II combined will result in $O(dn)$ inequalities. This is because, when an edge is processed, either the path to its neighbor is improved (Cat I), or it remains unchanged (Cat II), and hence every edge results in at least one inequality. Since the average degree per node is $d$, the resulting number of inequalities is $O(dn)$. The number of inequalities for Cat III is $O(n)$ since one inequality of Cat III is generated for every vertex processed. Thus, the complexity of the model is $O(dn)$. Since most large real graphs are sparse, i.e., $d \ll n$ (generally $d$ is of the order of tens or hundreds), we refer to this model as the *Linear* model with complexity growing linearly with $n$.

## 4.2 All Pairs Shortest Paths – Quadratic Model

In Theorem 2.1, we formalized the concept of *composability* of models formed using Anónimos. We now demonstrate composability by combining the models for *single source shortest paths tree* for modeling *all pairs shortest paths* problem [9]. All pairs shortest paths is a case where every vertex in the graph is considered as a *single source shortest paths tree*. Anónimos can also use Floyd-Warshall [10] algorithm for all-pairs shortest paths. We prefer to build on Dijkstra's algorithm since in addition to demonstrating *composability*, we can build on the models developed in the previous section, and Dijkstra's algorithm has additional properties (described in Section 6) which makes it better suited for certain applications.

A simple solution for the all-pairs problem is to generate the **Linear** model (as in Section 4.1) for the single source shortest paths trees for all the vertices $v_1, v_2, \ldots, v_n$, obtain the set of constraints $S_1, S_2, \ldots, S_n$, and then obtain the model for all-pairs as $S_1 \cup S_2 \cup \cdots \cup S_n$. Since each of the $S_i$'s provide constraints on *all* edges, hence the constraints from one *single source shortest paths tree* cannot contradict the assignment of another tree.

THEOREM 4.2. *A model comprised of all the constraints generated by the **Linear** solution for single source shortest paths tree, repeated for all the vertices of the tree, is a correct model for the all-pairs problem.*

PROOF. **Proof by Contradiction.** Let $G = (V, E, W)$ be the input graph, and $G' = (V, E, W')$ be the anonymized graph. Let us assume that there exists at least one pair of vertices $v_i, v_j$ whose shortest paths in $G'$ differs from its shortest path in $G$. The shortest path from $v_i$ to $v_j$ in the all-pairs problem is the path from $v_i$ to $v_j$ in the single source shortest paths tree with $v_i$ as the source, i.e., $T_i$. This implies that $T_i$ in $G$ does not match $T'_i$ in $G'$, which is a contradiction of Theorem 4.1. □

**Complexity of the Model.** The complexity of the model can be derived trivially from the complexity of the constituting model. Each of the shortest paths trees have a complexity of $O(dn)$, and this repeated for $n$ vertices gives us a total complexity of $O(dn^2)$.

## 5. OPTIMIZING THE MODELS

In the previous section, we developed a couple of simple models for the shortest paths problem, and demonstrated the composability of the models. We now provide optimizations to the simple models to reduce the complexity of the models while relaxing the composability property of the models – composability of the optimized models require special handling which we also discuss later in the section.

### 5.1 Single Source Shortest Paths – Reduced Model

We exploit specific properties of shortest paths to reduce the complexity of the naïve application of Anónimos to Dijkstra's algorithm which resulted in the **Linear** model. Note that even though Dijkstra's algorithm tries to relax the neighbors when processing a vertex, the ultimate goal is to select an appropriate vertex for the next iteration, i.e., the vertex with the smallest known cost from the source. Category III inequalities model this information in an efficient way, and hence ideally, only Category III inequalities are needed. However Category III inequalities only include edges that are part of the shortest paths tree. Therefore, if **only** Category III inequalities are considered in the model, then only part of the total number of edges are modeled. These inequalities by themselves do not put constraints on non-tree edges, and thus, if no care is taken while reassigning edge weights in the anonymized graph, it can lead to violations of the order in the anonymized graph. For instance, if edge $(u, v)$ is a non-tree edge, then a model using only Category III would not impose any constraint on $(u, v)$. Hence a reassignment of weights in the anonymized graph might assign the edge $(u, v)$ a weight such that Dijkstra's algorithm executing on the anonymized graph selects $(u, v)$ as a tree edge.

Therefore, to ensure correctness, the model must be augmented to make sure that the non-tree edges are not included in the tree when the algorithm executes on the anonymized graph. The following theorem formalizes this proposition.

THEOREM 5.1. *A model which ensures that $(i)$ the order of selection of vertices remains the same even after anonymization, and $(ii)$ non-tree edges in the original graph are not included in the tree constructed on the anonymized graph, will also ensure that the shortest paths tree in the original and anonymized graph are also same, i.e., the model is correct.*

PROOF. **Proof by Contradiction.**
Let $G = (V, E, W)$ be the input graph, and $G' = (V, E, W')$ be the anonymized graph. Let $T$ be the shortest paths tree starting at vertex $v_0$ in $G$ and $T'$ be the corresponding tree in $G'$. Let us assume that $T$ and $T'$ are different. Let $v$ be first vertex where $T$ and $T'$ differ, and let $u$ be its predecessor in $T$, and $u'$ in $T'$ such that $u \neq u'$. Then the following two possibilities arise:
**Case I:** The edge $(u, v) \in T$, and $(u', v) \notin T$. Now if $u'$ is the predecessor of $v$ in $T'$, then $(u', v) \in T'$. But this is a contradiction since $(ii)$ ensures that if $(u', v) \notin T \Rightarrow (u', v) \notin T'$.
**Case II:** Both edges $(u, v)$ and $(u', v)$ are in $T$. If $(u', v)$ is a directed edge, then this is not possible since vertex $v$ can have only one predecessor in $T$ which is $u$, and since $(u', v)$ is a directed edge towards $v$, it cannot be included in the path to some other vertex processed after $v$ leading to a contradiction to the condition $(i)$. □

**Augmenting the model – Complexity and Correctness.** Category III inequalities enforce condition $(i)$ of Theorem 5.1. A simple solution to ensure that condition $(ii)$ is also satisfied is to keep track of the edges not in the shortest path tree $(T_s)$ and when assigning weights to the anonymized graph, non-tree edges are assigned weights greater than the shortest path with the largest weight. This ensures that these edges are not picked as the shortest paths in the anonymized graph. Thus, Category III inequalities along with some additional information can model Dijkstra's algorithm, and the complexity of the modified model becomes $O(n)$ ($n - 1$ to be exact). Algorithm 5.1 provides the pseudocode for the **Reduced** model described in Section 5.1, while Algorithm 5.2 provides the code for reassigning edge weights in the anonymized graph obtained using the *Reduced* model. The asymptotic complexity of the models in this section and in Section 4.1 are the same: both grow linearly with $n$ (assuming that $d$ is a constant compared to $n$). But considering the fact that $d$ is generally of the order of 10 or 100 (as shown in our experiments using social network graphs), the model suggested in this section provides 1 to 2 orders of magnitude reduction in the number of inequalities.

### 5.2 All Pairs Shortest Paths – Optimized Model

We now develop an efficient model for the all pairs shortest paths problem. The model obtained by composition of the *Linear* model,

**Algorithm 5.1** Reduced model
1: /* Initialize similar to Dijkstra in Algorithm 4.2. */
2: $T \leftarrow \phi$ /* Set of edges in the Tree. */
3: **while** $Q \neq \phi$ **do**
4:    $u \leftarrow ExtractMin(Q)$
5:    $S \leftarrow S \cup \{u\}$
6:    **if** $(\Pi(u), u) \notin T$ **then**
7:      $T \leftarrow T \cup \{(\Pi(u), u)\}$
8:    **end if**
9:    **if** $u' \neq \phi$ **then**
10:      $AddConstraint(f(v_0, u') \leq f(v_0, u))$
11:    **end if**
12:    $u' \leftarrow u$
13:    **for** each vertex $v$ such that $(u, v) \in E$ and $v \notin S$ **do**
14:      **if** $D[v_0, v] > D[v_0, u] + w[u, v]$ **then**
15:         $D[v_0, v] \leftarrow D[v_0, u] + w[u, v]$
16:         $\Pi(v) \leftarrow u$ /* Shorter path exists. */
17:      **end if**
18:      **if** $v \notin Q$ **then**
19:         $Q \leftarrow Q \cup \{v\}$
20:      **end if**
21:    **end for**
22: **end while**

**Algorithm 5.2** Reassignment of weights in Reduced model
**Require:** $v_l$ is the last vertex processed by Algorithm 5.1
1: **for** each edge $(u, v) \in E$ **do**
2:    **if** $(u, v) \in T$ **then**
3:      $w'[u, v] \leftarrow$ Value obtained from solution of model.
4:    **else**
5:      $w'[u, v] \leftarrow D'[v_s, v_l] + rand()$ /* $v_s$ is the source vertex. */
6:    **end if**
7: **end for**

**Algorithm 5.3** Optimized model for *all pairs shortest paths*
1: Run Algorithm 5.1 for all vertices $v_1, \ldots, v_n$ to determine the trees $T_1, \ldots, T_n$ and models $S_1, \ldots, S_n$
2: $T \leftarrow \phi$
3: $S \leftarrow \phi$
4: **for** each $T_i$ in $\{T_1, \ldots, T_n\}$ **do**
5:    $S \leftarrow S \cup S_i$
6:    **for** each edge $(u, v) \in T_i$ **do**
7:      **for** each $T_k$ in $\{T_1, \ldots, T_n\}$ such that $(u, v) \notin T_k$ **do**
8:         $S \leftarrow S \cup \{f(v_k, v) < f(v_k, u) + x_{(u,v)}\}$
9:      **end for**
10:    **end for**
11:    $T \leftarrow T \cup T_i$
12: **end for**

though correct, has many redundant inequalities. For example, edges that are not part of any of the trees need not be part of the model, and can be treated as the non-tree edges in Section 5.1. However in the described model, there are no means for filtering out these inequalities. We now delve deeper into the problem and show how the *Reduced* model described in Section 5.1 can be composed for the all pairs problem.

Note that two *Reduced* models cannot be merged in a naïve way (refer to [8] for more details), since all edges in the graph are not part of the Reduced model (recall that only edges in the single source shortest path tree are part of model). When merging the constraints of multiple trees developed using the Reduced model, some edges that are part of some trees but not all the trees may result in problems. We formalize this as follows:

DEFINITION 5.1. ***Problematic edges:*** *An edge $(u, v)$ is said to be **problematic** for composition if there exists a shortest paths tree $T_i$ such that $(u, v) \in T_i$, and there exists a tree $T_j$ ($T_i \neq T_j$) such that $(u, v) \notin T_j$.*

A problematic edge $(u, v) \notin T_j$ will not have any constraint involving $x_{(u,v)}$ in the model developed for $T_j$, and hence the constraints of $T_i$ (or any other tree $T_k$ which contains $(u, v)$) can set a value $w'[u, v]$ in the anonymized graph such that when $T'_j$ is reconstructed in the anonymized graph, $(u, v)$ is selected as an edge in $T'_j$. There was in fact a decision which the algorithm took when $(u, v)$ was not included in $T_j$, but since $(u, v)$ was not selected in $T_j$, this decision was not part of the reduced model. We now devise a mechanism to model this decision in $T_j$, so that the edge will no longer be problematic for $T_j$.

PROPOSITION 5.2. ***Eliminating Problematic Edges:*** *A problematic edge $(u, v)$ is not selected in $T_j$, if there exists another path from the source vertex $v_j$ to $v$ which is cheaper than the path from $v_j$ to $v$ through the vertex $u$, i.e., $D[v_j, v] < D[v_j, u] + w[u, v]$. If the corresponding constraint $f(v_j, v) < f(v_j, u) + x_{([u,v])}$ is added to the model of $T_j$, then $(u, v)$ is no longer a problematic edge for $T_j$. Similarly, if the process is repeated for all trees $T_k$ such that $(u, v) \notin T_k$, then $(u, v)$ is no longer a problematic edge for any of the trees.*

Therefore, once we have ensured that the problematic edges are eliminated during the composition of the constraints of the individual trees, we can compose the individual constraints to form a correct model for all-pairs shortest paths. Therefore, if $T_1, \ldots, T_n$ are the trees and $S_1, \ldots, S_n$ are the corresponding set of constraints, then we want to form $S = S_1 \oplus S_2 \oplus \cdots \oplus S_n$ which would model the all-pairs shortest paths problem. The algorithm composes the trees one at a time. The pseudocode for the algorithm generating the model is shown in Algorithm 5.3. Edges which are not in any of the trees can be dealt in a manner similar to the approach used in Section 5.1.

THEOREM 5.3. *The model created by composing individual trees while eliminating **problematic edges** preserves all the trees $T_1, \ldots, T_n$, and hence correctly models all pairs shortest paths.*

PROOF. **Proof by Mathematical Induction.**
**Base Case.** At the beginning of the algorithm, $T = \phi$ and $S = \phi$. Hence it is true trivially.
**Inductive Case.** Let us assume that after iteration $i$, we have $T$ and set of constraints $S$ that preserves trees $T_1, \ldots, T_i$, and at iteration $i + 1$, we are adding the tree $T_{i+1}$. Let us assume that $(u, v)$ is a problematic edge. For every $T_k$ such that $(u, v) \notin T_k$ ($T_k \in \{T_1, \ldots, T_n\}$), means that Dijkstra's algorithm did not pick $(u, v)$ in $T_k$, and addition of the constraint in line 8 makes sure that Dijkstra's algorithm executing on the anonymized graph will not pick $(u, v)$ as an edge in $T'_k$. This property exists in the original graph that made sure that $(u, v)$ was not picked in any of $T_k$. Therefore, it is evident that when the edge $(u, v)$ is added, the algorithm makes sure that it is not problematic, and hence at the end of the iteration, the set of constraints $S$ preserves trees $T_1, \ldots, T_i, T_{i+1}$.

Therefore, by the principle of mathematical induction, the set of constraints at the end of the algorithm preserved the trees $T_1, \ldots, T_n$, and hence in the anonymized graph, all the trees can be reconstructed which are identical to the trees in the original graph. □

THEOREM 5.4. *A model that preserves the trees $T_1, \ldots, T_n$ correctly models the shortest path between all pairs of vertices.*

PROOF. **Proof by Contradiction.** Let $G = (V, E, W)$ be the input graph, and let $G' = (V, E, W')$ be the anonymized graph. Let us assume that there exists at least one pair of vertices $v_i, v_j$ whose shortest paths in $G'$ differs from its shortest path in $G$. The shortest path from $v_i$ to $v_j$ in the all-pairs problem is the path from $v_i$ to $v_j$ in the single source shortest paths tree with $v_i$ as the source, i.e., $T_i$. This implies that $T_i$ in $G$ does not match $T'_i$ in $G'$, which is a contradiction, since the $T_i$ is preserved by Theorem 5.3. □

**Complexity of the Model.** The analysis of the complexity of the algorithm is a bit more involved. In the *best case*, all the trees have the same edges. Since there are no problematic edges, no new constraints were added, and hence the complexity is $O(n^2)$. In the *worst case*, every problematic edge will add $O(n)$ inequalities, and again, there can be at most $O(dn)$ problematic edges. Therefore, the number of added constraints are:

$$\underbrace{(n-1) + (n-1) + \cdots + (n-1)}_{dn \text{ terms}}$$
$$+ \underbrace{(n-1) + (n-1) + \cdots + (n-1)}_{n \text{ terms}}$$
$$= O(dn^2) + O(n^2) = O(dn^2)$$

Therefore, the total number of inequalities is $O(dn^2)$. Thus the complexity is no worse than the model described in Section 4.2. Our experimental evaluation on real datasets shows that this model performs significantly better on the average than $O(dn^2)$.

## 6. APPLICATION SPECIFIC PROPERTIES

In addition to the properties described in Section 2.1 which are general to Anónimos, there are some additional properties which are interesting for the shortest paths problem. While generating the constraints for the model, the algorithm can be terminated at any point prior to completion, and this early termination has specific applications for the shortest paths problem. Formulating Anónimos as an LP problem ensures that the model being constructed is consistent at every point during the execution, and hence, these interesting sub-properties are also preserved. Furthermore, since the algorithm does not process every node and vertex in the graph, this might lead to considerable savings in the complexity of the resulting models.

**Early termination of Single Source Shortest Paths.** When computing the shortest paths tree from a specific node in the tree, the vertices are processed in the order of increasing distance from the source vertex. Therefore, if the algorithm is terminated after processing $k$ of the $n$ vertices, the resulting model preserves the $k$ Nearest Neighbors (kNN) of the starting node. kNN is often useful when the start vertex is an "influential user", and the kNN tree provides a path for information spread [14]. In the Reduced model (Section 5.1), an inequality is added for every vertex processed. Therefore, the complexity of the proposed model is $O(k)$ compared to $O(n)$ for the entire tree, a significant saving when $k \ll n$.

**Early termination of All Pairs Shortest Paths.** When combining individual shortest paths trees, the algorithm can be terminated after processing trees $T_1, \ldots, T_k$ (where $k < n$) and the model is still consistent for these trees and preserves the all pairs shortest paths for paths between vertices $1 \ldots k$. In addition, the trees $T_1, \ldots, T_k$ can be chosen as well as composed in any arbitrary order. Consider the *"Los Angeles"* community example in Section 1.

**Table 2: Summary of Complexity of the models.**

| Single source | All pairs |
|---|---|
| **Linear:** $O(dn)$ | **Quadratic:** $O(dn^2)$ |
| **Reduced:** $O(n)$ | **Optimized:** $O(n^2)$ (best), $O(dn^2)$ (average) |

**Table 3: Summary of the Social Graphs.**

| Data Set | No. of Vertices | No. of Edges | Avg. Degree |
|---|---|---|---|
| **Flickr-user-3** | 55,803 | 6,662,377 | 119.39 |
| **LJ-user-3** | 15,508 | 384,947 | 24.82 |
| **Orkut-user-3** | 26,110 | 899,638 | 34.46 |
| **Youtube-user-3** | 237,469 | 2,457,206 | 10.35 |
| **Flickr-comm** | 1,382 | 69,321 | 50.16 |
| **LJ-comm** | 1,497 | 21,481 | 14.35 |
| **Orkut-comm** | 1,047 | 28,240 | 26.97 |
| **Youtube-comm** | 1,823 | 29,342 | 16.1 |

A client requesting anonymized data corresponding to all the members in the *"Los Angeles"* community might only be interested in shortest paths between all pairs of "computer scientists". In such a scenario, only the shortest paths trees with "computer scientists" as roots need to be combined. If the number of trees $k \ll n$, then this technique will have a complexity of $O(kn)$, i.e., linear in the number of vertices in the graph.

## 7. EXPERIMENTAL EVALUATION

In this section, we experimentally evaluate the different models based on Anónimos, compare their complexity, and validate our analysis (Table 2 provides a summary of complexity of the models). All the algorithms were implemented in Java, and the experiments were run on a 2.4GHz Intel Core 2 Quad processor. The machine has 3GB main memory and runs Fedora Core Linux. We used four real social network data sets obtained from the authors of [22]. In our experiments, we used a free open-source LP Solver (*lp_solve 5.5*) [21]. We report the time taken to generate the model, complexity of the model, and the time taken to solve the models. We assume that the LP solver is de-coupled from the system generating the model. Therefore, the model is written to disk, and the system solving the model reads the model from disk, and generates the solution, which is then used to anonymize the model. Hence the reported times include the disk access latencies. Most open source implementations of LP solvers are not heavily optimized, and are stable for smaller systems. There are commercial systems which are much faster than these open source implementations, and can also handle larger models. Correctness of the models is also experimentally validated by checking the equivalence of the shortest paths trees and all-pairs shortest paths in the original and the anonymized graphs. As mentioned in Section 2.1.1, the choice of objective function provides some flexibility to the publisher. We experimented with a number of objective functions such as setting all coefficients to unity (unity objective function), or setting them to random values picked from uniform as well as gaussian distributions, but no significant difference in degree of anonymization was observed. In the reported experiments, we use a unity object function.

**Implementation Issues of Anónimos.** It is appropriate to address some of the subtleties of our implementation for completeness. Every modeled decision results in an inequality. In order to deal with ties and different implementations of queues or other structures used, the ties in the original graphs should be modeled exactly in the same way in which it was resolved while generating the model in the original graph. Consequently, if the edge $(u_1, v_1)$ was chosen by the algorithm ahead of edge $(u, v)$, to ensure that the algorithm takes the same decision even in the anonymized graph, we model the decision as $x(u_1, v_1) \leq x(u, v) - \epsilon$, where $\epsilon > 0$ is a small real number. Additionally, LP solvers do not accept strict inequalities of the type $x(u, v) < b$. Therefore, such inequalities are converted to non-strict inequalities as $x(u, v) \leq b - \epsilon$, where again $\epsilon > 0$ is a small real number.

## 7.1 Graph Data sets

Mislove et al. [22] crawled a number of social network sites for analyzing the properties of these large social graphs, and have made their data sets publicly available. Their data sets include the graphs for a number of popular social networking sites: Flickr (www.flickr.com), Live Journal (www.livejournal.com), Orkut (www.orkut.com), and Youtube (www.youtube.com). While Orkut is a pure social networking site, LiveJournal (referred to as **LJ** in the data sets) is a blogging site whose users form a social network, while Flickr and Youtube are photo sharing and video sharing sites respectively, with an overlayed social network structure amongst its users. We model the graphs of these networks as directed graphs where edges have positive weights, but the models can be extended for undirected graphs. The published graph data sets are unweighted, but since our model is not dependent on the semantics of the weights or their magnitude, we assign randomly generated weights (real numbers in the range 1 to 100) to the edges of the graph. We used different distributions for assigning edge weights, but no considerable change in complexity was observed.

The social network data sets have two specific forms of sub-graphs:

**User Driven Structures:** These are sub-graphs where a specific user is of interest, and is useful for applications focussed on a single user. For example, for marketing purposes, a company might select some *influential* users for free trials of their products so that they can influence other users to use or buy the product [14]. Similarly, applications such as *shortest paths trees* and *nearest neighbors* will also be interested in similar structures. To simulate these structures, we select a vertex in the graph as the root, and extract the graph induced by the vertices which are within $k$ *degrees of separation* from the root (a vertex $v$ is a first degree connection to the root $v_0$ if there exists an edge $(v_0, v)$). We use the *user* suffix to refer to the user data sets, and for our experiments, we consider $3^{rd}$ degree of separation (e.g., *Orkut-user-3*).

**Community Driven Structures:** These graphs correspond to communities (or groups) within the social networks. For example, in our examples in Section 1, we refer to the *"Los Angeles"* community in Facebook. Community structures are very important for applications such as *targeted advertising*, *shortest paths*, *nearest neighbors* etc. This is primarily since users in the same community share common interests, and hence many applications can be driven by the community structure. For the experiments, we select communities inside the social networks, and extract the graph induced by the members of the community. We use the *comm* suffix to refer to the community data sets (e.g., *Orkut-comm*).

Table 3 summarizes the different graphs in the data set used in our experiments in terms of the number of vertices, number of edges, and average out-degrees. To provide better insight into the distribution of the out-degrees of the vertices, in Figure 2, we plot the *cumulative distribution function (CDF)* of the out-degrees of the graphs in the data set. Along the $x$-axis is the out-degree, and along the $y$-axis is the fraction of the total number of vertices whose out-degree is less than the corresponding value of the $x$-axis. Figure 2(a) plots the CDF for the *user driven graphs*, while Figure 2(b) plots the CDF for the *community driven graphs*. Each line in the figure corresponds to a graph in the data set, and represents the fraction of vertices that have out degree less than or equal to the corresponding point on the $x$-axis. As can be noted from Figures 2(a) and 2(b), Flickr graphs have a considerably higher out degree compared to the other three graphs, and for the *user graphs*, about 12% of the vertices have an out degree higher than 250.

## 7.2 Single source shortest paths

In this section, we experimentally evaluate the models for *single source shortest paths tree* and demonstrate the benefits of our optimization. We compare the *Linear* model to the *Reduced* model in terms of the complexity of the model, and the time taken to build the model and write it to disk. Recall that the complexity of the model corresponds to the number of inequalities generated, and the time taken includes the time for executing Dijkstra's algorithm, generating the inequalities, and writing the generated inequalities to disk.

Figures 3 and 4 provide a comparison of the two modeling techniques for both types of data sets for all the social graphs. Figures 3(a) and 3(b) compare the complexity of the models, while Figures 4(a) and 4(b) compare the time taken to build the model. In all the figures, the $x$-axis represents the social graphs, and the $y$-axis for Figures 3(a) and 3(b) plots the number of inequalities constituting the model, while the $y$-axis for Figures 4(a) and 4(b) plot the time in seconds. Note that the $y$-axis of all the plots have been plotted in logarithmic scale. It is evident from the figures that the *Reduced* model is extremely efficient compared to the *Linear* model both in terms of complexity and time. The complexity of the *Reduced* model is about 1 to 2 orders of magnitude lesser when compared to the *Linear* model and so is the time taken in computing the model.

Table 4 provides the results from these experiments along with a detailed breakup of the number of inequalities, as well as the reduction in complexity and time of the *Reduced* model compared to the *Linear* model. For the *Linear* model, the categories of inequalities in Table 4 correspond to the categories defined in Section 4.1. As is evident from Table 4, the *Reduced* model provides about $O(d)$ times improvement in complexity of the models for all the graphs, as observed in Section 5.1. Depending on the graph, the value of $d$ varies, and so does the factor of improvement. For example, for the *Flickr-user-3* data set, $d$ is 119.39, and the complexity of the *Reduced* model is about 120 times less than that of the *Linear* model. The large reduction in the number of inequalities also affects the time to build the model, since in the *Linear* model, fewer inequalities need to be *generated*, and more importantly, fewer inequalities need to be *written to disk*. This is illustrated by the almost 90% improvement in time to generate the *Reduced* model.

## 7.3 All pairs shortest paths

In this section, we experimentally evaluate the models for the *all pairs shortest paths problem*. In a community of a social network, users share common interests, and an application that uses minimum cost paths between any two members of the community would require the all-pairs shortest paths. On the other hand, for a *user driven* social graph, two users in the graph might be completely unrelated, and from an application's perspective, shortest paths between them are not interesting. Thus, we evaluate the models for *all-pairs* only for the *community driven graphs*.

### 7.3.1 Evaluating shortest paths between all pairs

We refer to the model of Section 4.2 as the *Quadratic* model, and that of Section 5.2 as the *Optimized* model. Figure 5 compares the two models in terms of complexity and the time taken to build the model, and both of these terms have the same meaning as described in the previous section. In Figure 5(a), the number of inequalities in the models is plotted along the $y$-axis, while in Figure 5(b), the time taken (in seconds) to generate the model is plotted along the $y$-axis. In both the figures, the x-axis represents the different graph data sets, and again note that the $y$-axis is plotted in logarithmic scale. As noted in Section 5.2, Figure 5 illustrates the benefits of the *Optimized* model compared to the *Quadratic* model both in terms of complexity and time.

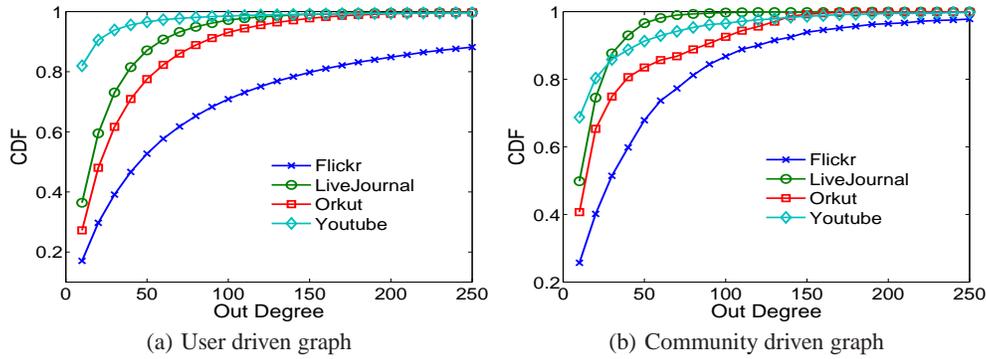

(a) User driven graph

(b) Community driven graph

**Figure 2: Cumulative Distribution Function for the out-degrees of the different graph data sets used for the experiments.**

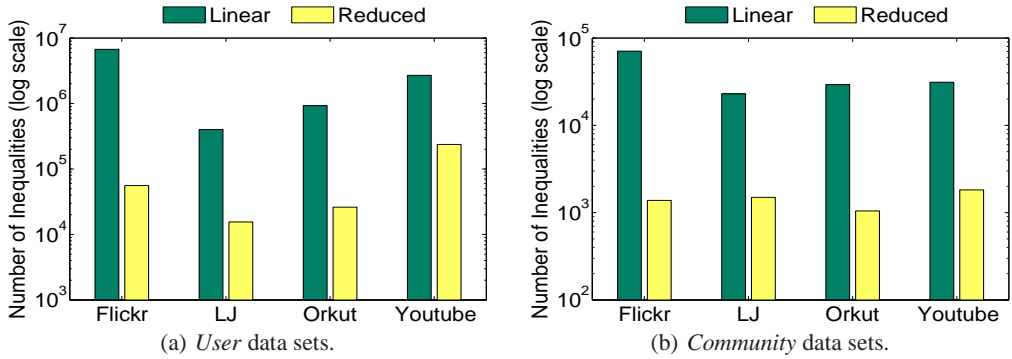

(a) *User* data sets.

(b) *Community* data sets.

**Figure 3: Complexity of the models for *single source shortest paths* tree.**

Table 5 provides the experimental results, tabulating the break down of the categories of the constituent inequalities that form the model. For the *Quadratic* model, the categories of the inequalities correspond to the ones defined in Section 4.1. For the *Optimized* model, the *Merge* inequalities are the ones generated when the individual shortest paths trees are merged into one consistent model compensating for the problematic edges, while the *Trees* inequalities are the total number of inequalities generated for the trees. Since this corresponds to *Category III* inequalities (as the *Reduced* model for single source only uses *Category III* inequalities), Columns 4 and 8 of Table 5 are identical. Table 5 also provides data that allows a deeper analysis of the reasons for the improved performance of the *Optimized* model, even though both models have the same complexity bound $O(dn^2)$. As was noted in Section 5.2, the inefficiency of the *Quadratic* model stems from the fact that it cannot leverage the absence of some edges from all the trees, which allows these edges to be excluded from the model. These edges are represented by the column titled *Unconstrained Edges* in Table 5. It can be seen that in all the social graphs, a high percentage of edges are not part of any tree, and eliminating these edges from the model considerably simplifies the model. This is evident from the 70–80% reduction in complexity of the *Optimized* model compared to the *Quadratic* model. As seen in the case of *single source shortest tree*, reduction in complexity of the model also considerably reduces the time, primarily because fewer inequalities are written to disk.

### 7.3.2 Benefits of Early Termination

We now experimentally evaluate the benefits of early termination of the all-pairs model as explained in Section 6. This kind of model finds application in *community driven* graphs where the application is interested in a specific subset of vertices. Our goal is to validate our analysis that if *all-pairs shortest paths* are not required by the application, we can considerably reduce the complexity of the model. For our experiments, we consider two cases, one where the pairwise shortest paths between a subset of 100 vertices is to be determined and the shortest paths tree for the rest suffices, and in another, we are interested in a subset of 200 vertices. These subsets are randomly selected, and simulate our example where the application is interested in "computer scientists" or "guitarists" in the *"Los Angeles"* community of Facebook. Table 6 summarizes the results from these experiments, and compares them with the numbers obtained for the *Optimized* solution for all-pairs shortest paths from Table 5. It is evident from the table that when all possible shortest paths need not be preserved, there can be a significant reduction in the complexity of the models. This significant reduction (reflected by the rightmost columns in Table 6) is obtained primarily due to the fact that a huge portion of the inequalities for the merge phase is not required for the trees that are not of interest and therefore do not need to be merged.

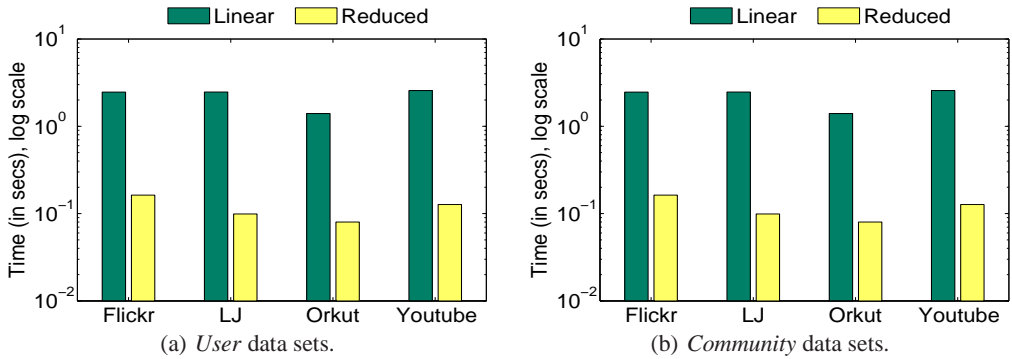

Figure 4: Time to build the model for the *single source shortest paths* tree.

Table 4: Experimental evaluation of single source shortest paths tree.

| Data Sets | Linear Model | | | | | Reduced Model | | Summary | |
|---|---|---|---|---|---|---|---|---|---|
| | Number Inequalities | | | | Time | Number of | Time | Times Reduction | % Reduction |
| | Cat I | Cat II | Cat III | Total | Taken (s) | Inequalities | Taken (s) | in Complexity | in Time |
| **Flickr-user-3** | 204,626 | 6,457,751 | 55,802 | 6,718,179 | 98.81 | 55,802 | 2.835 | 120.39 | 97.13 |
| **LJ-user-3** | 39,030 | 345,917 | 15,507 | 400,454 | 4.783 | 15,507 | 0.938 | 25.83 | 80.39 |
| **Orkut-user-3** | 72,130 | 827,508 | 26,109 | 925,747 | 15.735 | 26,109 | 1.752 | 35.47 | 88.87 |
| **Youtube-user-3** | 417,526 | 2,039,680 | 237,468 | 2,694,674 | 44.943 | 237,468 | 8.226 | 11.35 | 81.7 |
| **Flickr-comm** | 4,112 | 65,209 | 1,381 | 70,702 | 2.464 | 1,381 | 0.163 | 51.2 | 93.39 |
| **LJ-comm** | 3,148 | 18,333 | 1,496 | 22,977 | 2.471 | 1,496 | 0.099 | 15.36 | 95.99 |
| **Orkut-comm** | 2,409 | 25,831 | 1,046 | 29,286 | 1.401 | 1,046 | 0.08 | 27.99 | 94.29 |
| **Youtube-comm** | 3,605 | 25,737 | 1,822 | 31,164 | 2.564 | 1,822 | 0.127 | 17.11 | 95.05 |

## 7.4 Overall time overhead

In all the above experiments, we considered only the complexity of the model, and the time taken to generate the model. Once the model has been generated, it has to be solved to anonymize the graph. The time required for this step depends on the efficiency of the LP solver. We used an open source LP Solver [21] in our experiments, and it is widely acknowledged that commercial LP solvers are far more efficient compared to open source implementations. As example timings, for the *Reduced* model of *single source shortest paths tree* problem, the LP solver took 0.394 seconds to solve the model for *Orkut-comm* graph, 0.541 seconds for the *Youtube-comm* graph, 150.638 seconds for the *LJ-user-3* graph, and 629.869 seconds for the *Flickr-user-3* graph. For the all-pairs problem, where the complexity of the model rises to about $100K$ inequalities, the solvers took about an hour to find a solution. We remark that our open source LP solver is not optimized for solving large, sparse models, and these timings are not the best possible. Furthermore, solving the model constitutes an offline cost and hence the exact times are not significant for our evaluation.

## 7.5 Evaluating Data privacy

In this section, we evaluate the privacy preserving properties of the proposed models. In our evaluation, we use the two measures presented in Section 3, i.e., *k-anonymity* [26] and *Spearman rank correlation coefficient* [25]. As explained in Section 3, in the context of sensitivity of edge weights, both measures are defined in a neighborhood. Computation of *k-anonymity* of edges follow directly from its definition. The *Spearman rank correlation coefficient* is computed for every vertex in the graph. For each vertex in the original graph and the corresponding vertex in the anonymized graph, the list of edges emanating from the vertex comprises the ranked lists used for computing the coefficient. The lists are sorted by edge weights, and the coefficient measures the correlation between the ranks of the edges in the two lists. Figures 6, 7, and 8 provide the experimental results for the two measures on the real data sets. In these experiments, we use the *Reduced* model for *single source shortest paths* tree and the *Optimized* model for *all pairs shortest paths* problem.

Figures 6 and 7 plot the percentage of edges in the graph that are *k-anonymous* in their neighborhood for a given value of $k$ and indistinguishability threshold $\mu$. Figure 6 plots the graphs of the model for *single source shortest paths* tree and Figure 7 plots the graphs of the model for *all pairs shortest paths* problem. Along the $x$-axis we plot the different values of $k$, and along the $y$-axis, we plot the percentage of edges that are *k-anonymous* for the corresponding value of $k$ on the $x$-axis. Each graph plots two selected data sets and compares the *k-anonymity* of the original and anonymized versions of the same graphs. Different graphs correspond to different data sets, different values of $\mu$, and different algorithms. In these experiments, we selected the *Flickr* and *Orkut* graphs as representatives. Similarly, $\mu$ values of 1 and 3 are representatives chosen to show the variance of the anonymity levels as the indistinguishability threshold increases. In our experiments, the edge weights were in the range 1 to 100, so $\mu = 1$ corresponds to 1% of the total range of edge weights. As is evident from the Figures 6 and 7, our anonymization models considerably improve the *k-anonymity* of the anonymized graphs when compared to the original graphs. The improvement is even more significant for larger values of $k$ and smaller values of $\mu$, which demonstrates the improved anonymity of edges in the anonymized graph. Therefore, in the anonymized graphs, individual edge-weights are even less distinguishable. Note that this level of *k-anonymity* is provided by the model at no additional cost. We remark that *k-anonymity* can be further improved by adding constraints and setting bounds on the variables that ensure that the anonymized weights are even closer to each other. Additionally, note that the *k-anonymity* of the edges is better for the *Flickr* data set due to the higher average out-degree of the vertices which allows for more room for hiding in the *edge neighborhood*.

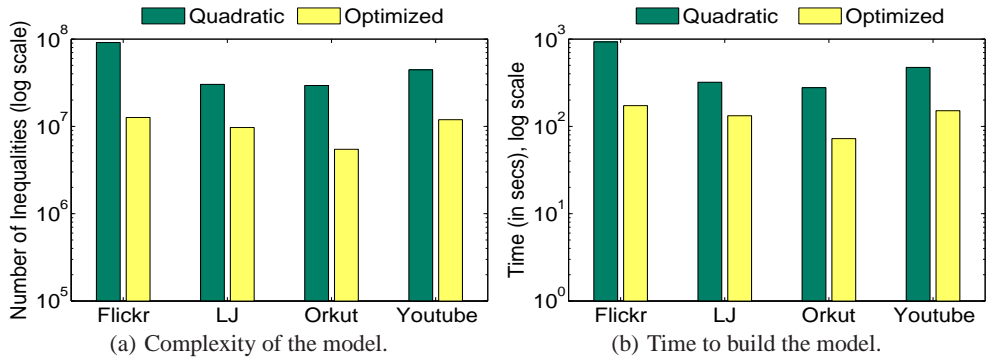

Figure 5: Performance of the models for the *all pairs problem*.

Table 5: Experimental evaluation of all pairs shortest paths problem for the community driven data sets.

| Data Sets | Quadratic Model | | | | | Optimized Model | | | | |
|---|---|---|---|---|---|---|---|---|---|---|
| | Number Inequalities | | | | Time Taken (s) | Number of Inequalities | | | Time Taken (s) | Unconstra- ined Edges |
| | Cat I | Cat II | Cat III | Total | | Merge | Trees | Total | | |
| Flickr | 3,645,749 | 85,824,651 | 1,813,512 | 91,283,912 | 926.71 | 10,837,381 | 1,813,512 | 12,650,893 | 172.66 | 60,166 |
| LJ | 2,330,938 | 25,847,924 | 2,107,957 | 30,286,819 | 320.42 | 7,588,195 | 2,107,957 | 9,696,152 | 132.31 | 15,003 |
| Orkut | 1,428,809 | 26,907,339 | 1,088,890 | 29,425,038 | 277.33 | 4,377,502 | 1,088,890 | 5,466,392 | 72.32 | 23,018 |
| Youtube | 2,762,305 | 38,902,975 | 2,756,994 | 44,422,274 | 473.945 | 9,163,912 | 2,756,994 | 11,920,906 | 151.04 | 22,802 |

Figure 8 plots the *Spearman rank correlation coefficient* of the models for *single source shortest paths tree* and *all pairs shortest paths* problem. Since the value of the coefficient $\rho$ forms a continuum in the range $-1.0 \leq \rho \leq 1.0$, for ease of presentation, we maintain a equi-width histogram of the coefficient values. Along the $x$-axis, we plot the bucket boundaries of the histogram, and along the $y$-axis we plot the percentage of vertices that have the value of $\rho$ in the range corresponding to the bucket. The two graphs plot four data sets and Figure 8(a) plots the results for the *single source shortest paths* tree while Figure 8(b) plots the results for *all pairs shortest paths* problem. Figures 8(a) and 8(b) demonstrate the excellent scrambling of the order of the edge weights. Note that $\rho = 0$ corresponds to no correlation of ordering, and the closer it is to 0, the harder it is for an adversary to determine the original order with high confidence. Our experiments show that for all data sets, more than 75% of vertices have $-0.3 \leq \rho \leq 0.3$, and about 90% of the vertices have $-0.5 \leq \rho \leq 0.5$. Additionally, note that the higher the average out degree (refer to Table 3 for the average degrees of the graphs in the data sets), the lesser the correlation between the original and the anonymized orders.

In conclusion, these experiments demonstrate the robustness of the privacy models, and show how hard it is for an adversary to determine the original edge weight, to uniquely identify edge weights, or to determine the original ordering of the weights, thereby effectively preserving the sensitivity of the weights.

## 8. RELATED WORK

The need to protect the privacy of social entities involved in social networks has given rise to active research in anonymization techniques for social network graphs. This interest has been primarily driven by the findings of Backstrom et al. [3] and Korolova et al. [15]. Backstrom et al. [3] described a technique based on the structural properties of graphs such as isomorphism and automorphism to re-identify vertices in the anonymized graph. Their technique was based on implanting unique structures in the graph which can be re-identified in the anonymized graph with very high probability. On the other hand, Korolova et al. [15] devised an attack where a node can be re-identified based in part on background information regarding the neighborhood. As a result, a lot of research has focused on *node identity anonymization* and *structural anonymization*. A comprehensive survey about the various anonymization techniques is provided in [18, 29].

A class of proposals, by Hay et al. [12], Zhou et al. [32], Liu et al. [19], and Zou et al. [33], suggest different methods for anonymization that are based on the addition and/or deletion of edges in the graph for altering the structure of the graph and the prevention of re-identification in the anonymized graph. On the other hand, Cormode et al. [7] suggest a technique for the anonymization of bipartite graphs based on safe groupings, an extension of which is class based anonymization [6]. Ying et al. [30] propose a randomization based spectrum preserving approach which effectively preserves the properties of the eigenvalues of the network, while anonymizing the edges, and Campan et al. [5] suggest a clustering based approach for node anonymization. Along different lines, Zheleva et al. [31] formulate the problem of edge re-identification in an unweighted graph, where the edge labels are sensitive.

The majority of existing work considers unweighted graphs for node identity and structural anonymization. But as reflected by recent work [17, 28], the weighted social network model is gaining importance, and so is edge weight anonymization. Liu et al. [20] suggest a probabilistic technique for anonymizing edge weights by perturbing the actual edge weights by a small $\sigma$ obtained from a probability distribution. The goal is to keep the total cost of the shortest path close to the cost of the path in the original graph. However in this approach, the anonymized weights are close to the original edge weights, and hence may reveal sensitive information about the original values. Anónimos aims at preserving general linear properties of the graph. For the shortest paths, our goal is to preserve the paths rather than the values and for most applications, the ability to reconstruct the actual path is more important than maintaining approximate values. In addition, if necessary, our model can approximately preserve the cost of the shortest paths as well by adding constraints of the form $f(u,\ldots,v) = D[u,v] \pm \epsilon$. Note that since the edge weights are only perturbed by a small value, the

Table 6: Experimental evaluation of all pairs shortest paths between a subset of vertices for the community driven data sets.

| Data Sets | 100 vertices | | 200 vertices | | Optimized All Pairs | | Percent Reduction | |
|---|---|---|---|---|---|---|---|---|
| | Number of Inequalities | Unconstrained Edges | Number of Inequalities | Unconstrained Edges | No. of inequalities | Unconstrained edges | 100 vertices | 200 vertices |
| **Flickr-comm** | 513,414 | 64,186 | 1,177,428 | 63,433 | 12,650,893 | 60,166 | 95.94 | 90.69 |
| **LJ-comm** | 314,107 | 18,339 | 732,212 | 17,819 | 9,696,152 | 15,003 | 96.76 | 92.45 |
| **Orkut-comm** | 253,002 | 25,709 | 562,005 | 25,429 | 5,466,392 | 23,018 | 95.37 | 89.72 |
| **Youtube-comm** | 374,516 | 25,596 | 835,831 | 25,162 | 11,920,906 | 22,802 | 96.86 | 92.99 |

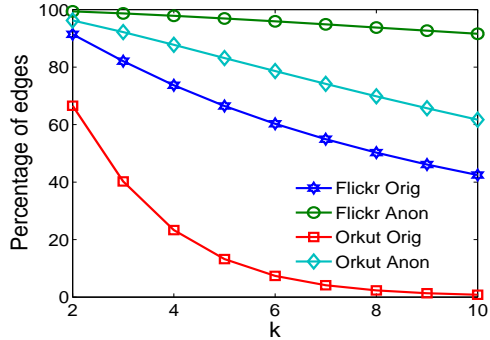
(a) *k anonymity* for single source shortest paths tree for user data sets and $\mu = 1$

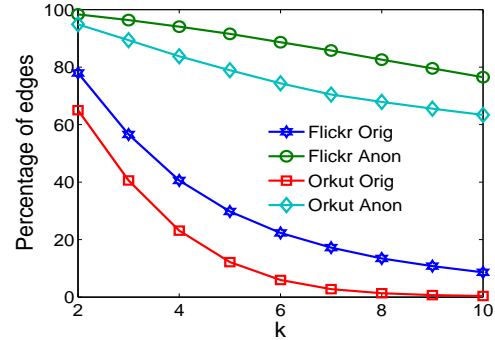
(b) *k anonymity* for single source shortest paths tree for community data sets and $\mu = 1$

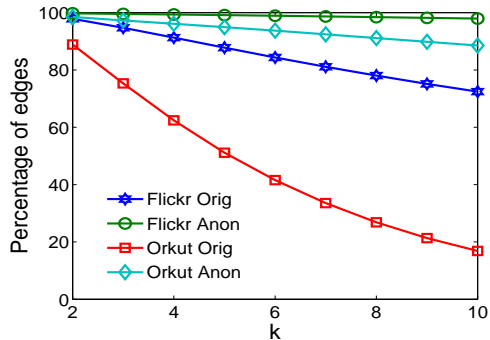
(c) *k anonymity* for single source shortest paths tree for user data sets and $\mu = 3$

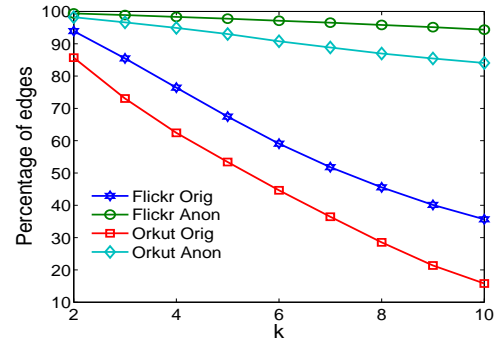
(d) *k anonymity* for single source shortest paths tree for community data sets and $\mu = 3$

Figure 6: Evaluating *k-anonymity* for *single source shortest paths* tree model.

technique of [20] can neither significantly improve *k-anonymity*, nor can it scramble the ordering of edge weights.

## 9. CONCLUSION

Anonymization of edge weights in a social network graph is important for enabling the analysis and mining of social graphs by computer scientists as well as social scientists. Such mining has significant impact on the management of social networks as well as the understanding of various social behaviors. We proposed Anónimos, a technique for the effective anonymization of weighted social network graphs by modeling *linear properties* and formulating them as an LP problem. The Anónimos approach is fairly straightforward and can be applied to preserve any *linear property* by a simple generation of inequalities corresponding to decisions made by the algorithm during its execution. As a proof of concept, we considered the *shortest paths problem* and showed how off-the-shelf LP packages can be used to effectively anonymize the graphs. The *composability* of Anónimos for preserving multiple properties in a single anonymized graph was demonstrated using the *all pairs shortest paths* problem. We also showed how a careful analysis of the properties can result in significant reductions in complexity of the models. Our experiments demonstrated the effectiveness of our techniques in the anonymization of graphs, and the efficacy of our optimizations, while ensuring that the desired properties are preserved across anonymization. In the future, we would like to explore extensions of Anónimos for other applications such as graph clustering, information spread modeling, etc., which also rely on linear combinations of edge weights.

## Acknowledgement

The authors would like to thank Divyakant Agrawal, Pamela Bhattacharya, and Sayan Ranu for their insightful comments on the earlier versions of the paper which has helped in improving this paper. The authors would also like to thanks Alan Mislove for providing the data sets used for the experiments. This work is partially supported by NSF Grants IIS-0744539 and IIS-0847925.

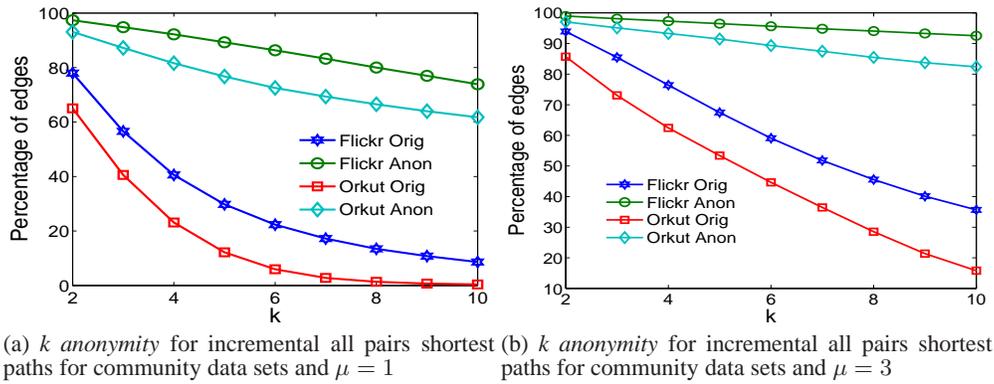

(a) *k anonymity* for incremental all pairs shortest paths for community data sets and $\mu = 1$

(b) *k anonymity* for incremental all pairs shortest paths for community data sets and $\mu = 3$

**Figure 7: Evaluating *k-anonymity* for the *all-pairs shortest paths* tree models. For incremental all pairs, algorithm is terminated after 100 vertices.**

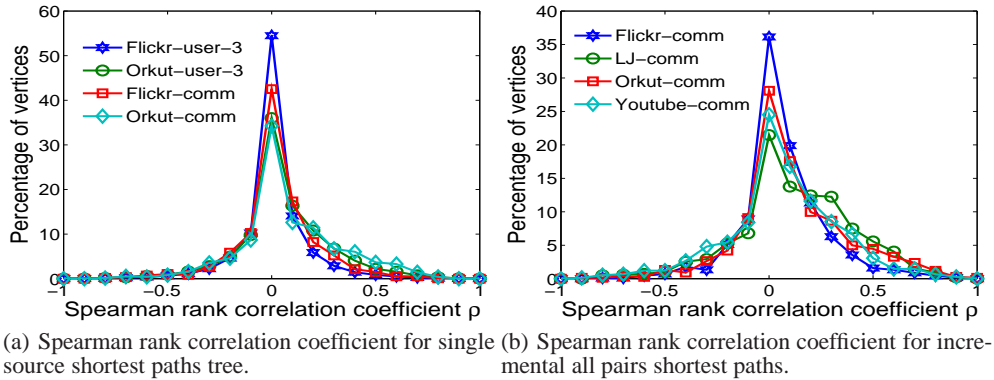

(a) Spearman rank correlation coefficient for single source shortest paths tree.

(b) Spearman rank correlation coefficient for incremental all pairs shortest paths.

**Figure 8: Evaluating *Spearman rank correlation coefficient* for the models. For incremental all pairs, algorithm is terminated after 100 vertices.**